\documentclass[%
 reprint,
 superscriptaddress,
 amsmath,amssymb,
 aps,
 prx,
floatfix,
]{revtex4-2}
\usepackage[english]{babel}
\usepackage{graphicx}
\usepackage{dcolumn}
\usepackage{bm}
\usepackage{comment}
\usepackage{xcolor}
\usepackage{amsmath}
\usepackage{braket}
\usepackage[normalem]{ulem}
\usepackage{float}
\usepackage[colorlinks,citecolor=red,urlcolor=blue,bookmarks=false,hypertexnames=true]{hyperref}

\usepackage[autostyle]{csquotes}
\MakeOuterQuote{"}

\usepackage{todonotes}

\begin{document}

\renewcommand{\d}{{\rm d}}

\preprint{APS/123-QED}

\title{Enhanced Photon Routing Beyond the Blockade Limit Via Linear Optics}

\author{Harjot Singh}
\altaffiliation{These authors contributed equally to this work}
 \author{Jasvith Raj Basani}
 \altaffiliation{These authors contributed equally to this work}
 \author{Edo Waks}
 
\affiliation{%
Institute for Research in Electronics and Applied Physics, and Joint Quantum Institute, Department of Electrical and Computer Engineering, University of Maryland, College Park, MA, USA 
}%

\date{\today}

\begin{abstract}
Directing indistinguishable photons from one input port into separate output ports is a fundamental operation in quantum information processing. The simplest scheme for achieving routing beyond random chance uses the photon blockade effect of a two-level emitter. But this approach is limited by a time-energy uncertainty relation. We show that a linear optical unitary transformation applied after the atom enables splitting efficiencies that exceed this time-energy limit. We show that the linear optical unitary improves the splitting efficiency from 67\% to 82\% for unentangled photon inputs, and from 77\% to 90\% for entangled photon inputs. We then optimize the temporal mode profile of the entangled photon wavefunction to attain the optimal splitting efficiency of 92\%, a significant improvement over previous limits derived using a two-level atom alone. These results provide a path towards optimizing single photon nonlinearities and engineering programmable and robust photon-photon interactions for practical,  high-fidelity quantum operations.
\end{abstract}

\maketitle


\section{\label{sec_1} Introduction}

Photonics provides a promising approach to realize scalable quantum technology~\cite{aspuru2012photonic,wang2020integrated,chang2014quantum,uppu2021quantum,elshaari2020hybrid}. Photons interact weakly with their environment, which makes them ideal carriers of quantum information~\cite{carolan2015universal,sparrow2018simulating,bunandar2018metropolitan,qiang2018large}.  At the same time, the lack of strong interactions poses a significant challenge to realize quantum operations between photons, which are essential for many quantum information processing applications. 

Single photon interactions require a strong optical nonlinearity. Bulk optical nonlinearities are an attractive option due to their potential for room temperature operation, but at this stage are still too weak~\cite{choi2017self, heuck2020photon, krastanov2021room}.  Alternatively, one can achieve nonlinearities at the single photon level by using a two-level emitter coupled to a cavity or a waveguide~\cite{kimble1998strong, javadi2015single, jeannic2022dynamical}. The nonlinear response of two-level emitters has been well studied~\cite{fan2010input, xu2015input, rephaeli2013dissipation} and has been realized experimentally using quantum dots~\cite{englund2007controlling, hennessy2007quantum}, atoms~\cite{birnbaum2005photon, rauschenbeutel1999coherent}, ions~\cite{takahashi2020strong} and superconducting circuitry~\cite{wallraff2004strong}. But the interactions mediated by a two-level emitter suffer from a time-bandwidth trade-off which limits the fidelity of operations such as the \texttt{CPHASE} gate~\cite{nysteen2017limitations}, photon sorting~\cite{ralph2015photon, yang2022deterministic} and photon routing~\cite{rosenblum2011photon}.

One particular application of interest is photon routing, where a two level emitter splits two indistinguishable photons into distinct output channels~\cite{aoki2009efficient}. Linear optical unitaries~\cite{reck,clements,scf} have been shown to reach peak splitting efficiencies of only 50\%~\cite{knill2001scheme,kok2007linear}. A two-level emitter exceed this limit, but cannot achieve perfect routing due to the time-bandwidth trade-off. An extensive analysis for the routing of two-photons has been performed by Rosenblum et \textit{al}. ref.~\cite{rosenblum2011photon}, where peak splitting efficiencies of 64\% and 68\% were attained for pulses with Lorentzian and Gaussian spectral profiles respectively. Engineering the time-energy relations by adding entanglement between the input photons further improves this efficiency to 77\% for an entangled pulses generated by a three-level atomic cascade emission~\cite{rosenblum2011photon}.

We show that the blockade limited splitting efficiency can be exceeded with the use of a linear optical unitary transformation after the atom. We optimize the unitary to achieve the best splitting efficiency for an uncorrelated two-photon input and show that it can exceed 82\% for a Gaussian pulse shape. We subsequently show that time-energy entangled inputs can achieve splitting efficiencies exceeding 90\%. Finally, we optimize the entangled photon wavefunction to achieve an optimal splitting efficiency of 92\%. This efficiency is significantly larger than the limit set by a two-level emitter alone with no unitary. In all cases, the unitary transformation fundamentally  changes the time-bandwidth trade-off, resulting in optimal performance at a reduced bandwidth of the input pulse compared to the bare two-level emitter.   

This manuscript is organized as follows: In section~\ref{sec_2} we derive the most general time-domain solutions for the probabilities of scattering event for two-photon wave packets incident on the two-level emitter and Mach-Zehnder Interferometer system. In section~\ref{sec_3} we find the optimal unitaries that maximize the splitting efficiency for both entangled and unentangled photons. In section~\ref{sec_4}, we additionally optimize the temporal wavefunction of the entangled photon input to achieve a more optimal splitting efficiency than would be possible by standard exponential or Gaussian temporal modes. Finally, section~\ref{sec_5} concludes the paper with a further discussion of the scope and impact of our work.


\section{\label{sec_2} System Model and Method}

\begin{figure}
    \centering
    \includegraphics[width = \columnwidth]{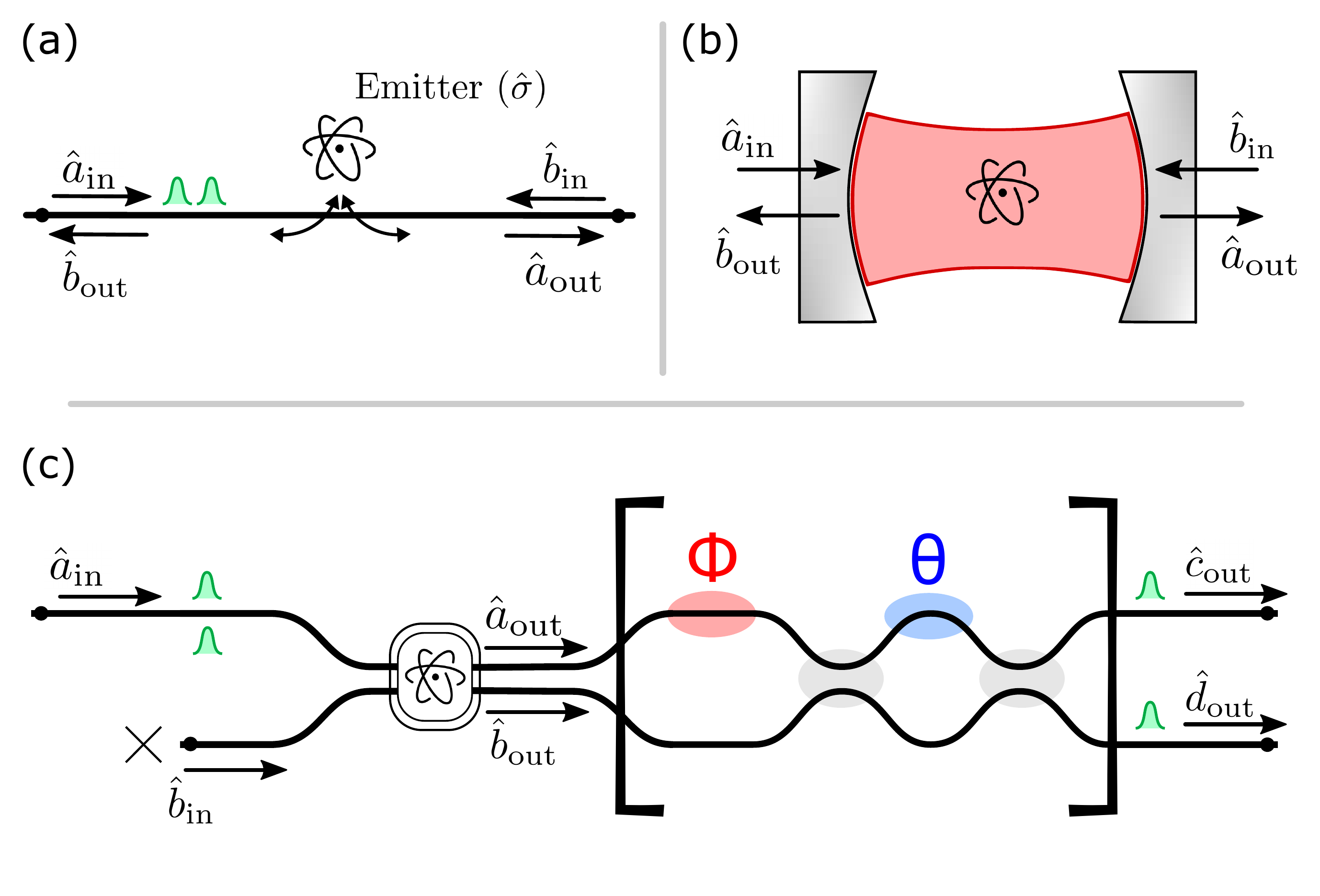}
    \caption{\textbf{Schematic representation of the photon splitting setup} \textbf{(a)} Schematic of two-level emitter coupled to a waveguide, indicating the directions of the input and output modes. \textbf{(b)} Alternative implementation using an atom coupled to a double-sided cavity. \textbf{(c)} Schematic for the photon splitting setup, where the output modes of the two-level emitter, coupled to a waveguide or cavity, are injected into a Mach-Zehnder Interferometer.}
    \label{fig:schematic}
\end{figure}

Fig.~\ref{fig:schematic}(a) shows the standard approach to single photon routing using a two-level emitter. The system is composed of a two-level emitter coupled to a waveguide.  The modes $\hat{a}_\mathrm{in}$ and $\hat{b}_\mathrm{in}$ are inputs to the emitter and $\hat{a}_\mathrm{out}$ and $\hat{b}_\mathrm{out}$ are the output modes. In the photon routing scenario, two photons are injected from mode $\hat{a}_\mathrm{in}$ and scatter into the two output modes.  Mode $\hat{b}_\mathrm{in}$ is in the vacuum state. Fig.~\ref{fig:schematic}(b) shows another way to implement this system, where a two-level emitter is coupled to a double-sided cavity. These two systems are equivalent in the bad cavity limit ($\gamma\ll\kappa$) where $\gamma$ represents the two-level emitter's spontaneous emission rate and $\kappa$ denotes the cavity decay rate.  In this limit, the cavity atom system can be replaced by a one dimensional atom model with a modified spontaneous emission rate given by $\Gamma=4g^2/\kappa$~\cite{scully1999quantum}, where $g$ is the atom-cavity coupling strength. 

Due to  photon blockade, the two input photons may be routed to spatially distinguishable output modes $\hat{a}_\mathrm{out}$ and $\hat{b}_\mathrm{out}$, an effect which we refer to as photon splitting. We define the photon splitting efficiency as the probability that two photons in the same input port exit at different output ports. Rosenblum et \textit{al.}~\cite{rosenblum2011photon}. extensively analyzed the splitting efficiency of a single atom and showed it was limited to 77\% due to a time-bandwidth tradeoff ~\cite{rosenblum2011photon}.

To improve the splitting efficiency, we consider the system in Fig.~\ref{fig:schematic}(c). We place a Mach-Zehnder Interferometer after the atom which applies a general linear optical unitary transformation given by: 
\begin{equation}
	\begin{bmatrix} \hat{c}_\mathrm{out} \\  \hat{d}_\mathrm{out} \end{bmatrix} = \begin{bmatrix} e^{i\phi} \sin(\theta/2) & \cos(\theta/2) \\ e^{i\phi} \cos(\theta/2) & -\sin(\theta/2) \end{bmatrix} \begin{bmatrix} \hat{a}_\mathrm{out} \\  \hat{b}_\mathrm{out} \end{bmatrix} 
	  \label{eq:MZI}
\end{equation}
In the above equations, $\hat{c}_\mathrm{out}$ and $\hat{d}_\mathrm{out}$ are the output modes of the unitary, which are directly related to the input modes via a scattering matrix. The scattering matrix has two tunable parameters, $\theta$ and $\phi$, which are represent applied phase shifts as shown in the figure. By tuning these two parameters we can implement any desired two-mode unitary transformation. We will use these two parameters to optimize the splitting efficiency into the output modes.

To calculate the splitting efficiency after the interferometer, we first define the time-ordered second order correlation functions:

\begin{equation}
\Gamma^{\mathrm{pq}}(\tau_1,\tau_2)  = \bra{\psi_o}\hat{p}^\dagger_{\mathrm{out}}(\tau_1)\hat{q}^\dagger_{\mathrm{out}}(\tau_2)\hat{q}_{\mathrm{out}}(\tau_2)\hat{p}_{\mathrm{out}}(\tau_1)\ket{\psi_o} 
\label{eq:gamma_pq}
\end{equation}
where $\{\hat{p},\hat{q}\} \in \{\hat{c},\hat{d}\}$. These correlations represent the probability densities that a photon is detected at time $\tau_1$, and a second photon is detected at time $\tau_2$. The wavefunction $|\psi_0\rangle$ represents the initial state of the system, which is assumed to be in the subspace where both photons are in mode $\hat{a}_\mathrm{in}$ and the atom is in the ground state. Because these are time-ordered correlations it is implicit that $\tau_2 \geq \tau_1$ in all calculations. Because we are restricting our attention only to a two photon input, the correlations can be written as $\Gamma^{\mathrm{pq}}(\tau_1,\tau_2)=|\psi_{\mathrm{pq}}(\tau_1,\tau_2)|^2$ where $\psi_{\mathrm{pq}}(\tau_1,\tau_2)$ is the correlation amplitude given by:
\begin{equation}
\psi_{\mathrm{pq}}(\tau_1,\tau_2)=\bra{0}\hat{q}_\mathrm{out}(\tau_2) \hat{p}_\mathrm{out}(\tau_1)\ket{\psi_0}
\label{eq:psi_pq}
\end{equation}

From these correlations, we can directly calculate the splitting efficiency $P_S$ as 
\begin{equation}
P_{S} = \int \d\tau_{1} \int \d\tau_{2} \left[ \Gamma^{\mathrm{cd}} (\tau_{1}, \tau_{2}) + \Gamma^{\mathrm{dc}} (\tau_{1}, \tau_{2}) \right]
\label{eq:p_s}
\end{equation}

To calculate the splitting efficiency, we apply the unitary to derive a relation between the correlation $\psi_{\mathrm{cd}}$ and $\psi_{\mathrm{dc}}$ in terms of the correlation amplitudes of the outputs of the atom given by 
\begin{equation}
\psi_{\mathrm{lm}} = \bra{0}\hat{m}_\mathrm{out}(\tau_2)\hat{l}_\mathrm{out}(\tau_1)\ket{\psi_0} 
\label{eq:psi_lm}
\end{equation}
where in the above $\{\hat{l},\hat{m}\} \in \{\hat{a},\hat{b}\}$. The above amplitudes are related to the output amplitudes of the interferometer via the relations
\begin{subequations}
\begin{equation}
\begin{aligned}
     \psi_{\mathrm{cd}} & = e^{-2 i \phi } \sin \left( \frac{\theta}{2} \right)  \cos \left( \frac{\theta}{2} \right) \psi_\mathrm{aa} - e^{-i \phi }  \sin ^2\left( \frac{\theta}{2} \right)\psi_\mathrm{ab} \\ & + e^{-i \phi } \cos ^2\left( \frac{\theta}{2} \right)\psi_\mathrm{ba} -  \sin \left( \frac{\theta}{2} \right) \cos \left( \frac{\theta}{2} \right) \psi_\mathrm{bb}
\end{aligned}   
\end{equation}
\begin{equation}
\begin{aligned}
     \psi_{\mathrm{dc}} & = e^{-2 i \phi } \sin \left( \frac{\theta}{2} \right)  \cos \left( \frac{\theta}{2} \right) \psi_\mathrm{aa} + e^{-i \phi }  \cos ^2\left( \frac{\theta}{2} \right)\psi_\mathrm{ab} \\ & -e^{-i \phi }  \sin ^2\left( \frac{\theta}{2} \right)\psi_\mathrm{ba} -  \sin \left( \frac{\theta}{2} \right) \cos \left( \frac{\theta}{2} \right) \psi_\mathrm{bb}
\end{aligned}   
\end{equation}
\label{eq:psi_cd_dc}
\end{subequations}
The above expressions enables us to directly calculate the output correlations of the interferometer from the correlation amplitudes of the atomic output modes. 

To calculate the correlation amplitudes of the atomic output modes, we use the standard Hamiltonian for the interaction of an atom with a waveguide given by $H=H_0+H_{\mathrm{int}}$ where~\cite{fan2010input, hofmann2003entanglement}:
\begin{equation}
\hat{H}_0= \int_{-\infty}^{\infty} \d\omega\, \omega(\hat{a}_{\omega}^\dagger\hat{a}_{\omega} +\hat{b}_{\omega}^\dagger\hat{b}_{\omega})
\label{eq:ham_wg}
\end{equation}
and
\begin{equation}
\hat{H}_{\mathrm{int}} = -i\sqrt{\frac{\gamma}{\pi}}\int_{-\infty}^{\infty} \d\omega\, [\hat{\sigma}^\dagger(\hat{a}_{\omega} + \hat{b}_{\omega})-\hat{\sigma}(\hat{a}_{\omega}^\dagger + \hat{b}_{\omega}^\dagger)]
\label{eq:ham_int}
\end{equation}
In the above equations, $\hat{\sigma}$ is the atomic lowering operator and $a_{\omega}$ are the bosonic reservoir mode operators for the waveguide.  They are related to the input modes via the relation $\hat{a}_{\mathrm{in}}(\omega)=1/\sqrt{2\pi}\smallint \d t ~ a_{\mathrm{in}}(t)e^{-i\omega t}$. The input and output modes are also directly related to each other by the input-output relations $\hat{a}_{\mathrm{out}}=\hat{a}_{\mathrm{in}}-\sqrt{2\gamma}\hat{\sigma}$ and $\hat{b}_{\mathrm{out}}=\hat{b}_{\mathrm{in}}-\sqrt{2\gamma}\hat{\sigma}$~\cite{fan2010input}.

As we feed the input pulse through $\hat{a}_{\mathrm{in}}$, $\hat{b}_{\mathrm{in}}$ is only a vacuum noise input. Since, we are calculating normally ordered moments of output operators, vacuum noise inputs from both input ports can be ignored and we can rewrite $\hat{b}_{\mathrm{out}}=\sqrt{2\gamma}\hat{\sigma}$. The initial state of the two photons in the input channel $\mathrm{a}_\mathrm{in}$ can be written as:
\begin{equation}
    \ket{\psi_{0}}=\int_{-\infty}^{\infty} \d t_{1}\int_{t_{1}}^{\infty} \d t_{2} ~ \xi(t_{1},t_{2}) \hat{a}^\dagger_{\mathrm{in}}(t_{1})\hat{a}^\dagger_{\mathrm{in}}(t_{2})\ket{0} \ket{g}
    \label{eq:init_state}
\end{equation}


In Appendix A, we show that the output correlation amplitudes after the atom are given by: 
\begin{widetext}
\begin{subequations}
\begin{equation}
   \psi_{\mathrm{bb}}(\tau_1,\tau_2)=4\gamma^2e^{-2\gamma(\tau_1+\tau_2)}\Biggl( \int_{\tau_1}^{\tau_2} \d t_2 \int_{-\infty}^{\tau_1} \d t_1 ~e^{2\gamma(t_1+t_2)} \xi(t_1,t_2) \Biggl)
\end{equation}

\begin{equation}
    \psi_{\mathrm{ba}}(\tau_1,\tau_2)=-2\gamma e^{-2\gamma \tau_1}\Biggl( \int_{-\infty}^{\tau_1} \d t_1  ~e^{2\gamma t_1}\xi(t_1,\tau_2) \Biggl) + \psi_{\mathrm{bb}}  (\tau_1,\tau_2)
\end{equation}

\begin{equation}
    \psi_{\mathrm{ab}}(\tau_1,\tau_2)= -2\gamma e^{-2\gamma \tau_2}\Biggl(\int_{-\infty}^{\tau_1} \d t_1 ~ e^{2\gamma t_1}\xi(t_1,\tau_1)  + \int_{\tau_1}^{\tau_2} \d t_2 ~ e^{2\gamma t_2}\xi(\tau_1,t_2) \Biggl) + \psi_{\mathrm{bb}} (\tau_1,\tau_2) 
\end{equation}
    
\begin{equation}
    \psi_{\mathrm{aa}} = \xi(\tau_{1}, \tau_{2}) + \psi_{\mathrm{ab}}(\tau_{1}, \tau_{2}) + \psi_{\mathrm{ba}}(\tau_{1}, \tau_{2}) - \psi_{\mathrm{bb}}(\tau_{1}, \tau_{2})
\end{equation}
\label{eq:scat_amps}
\end{subequations}
\end{widetext}

With these expressions for the two photon correlation amplitudes in the output modes of the emitter, we can directly calculate the photon splitting efficiency in eqn.~\eqref{eq:psi_cd_dc} and eqn.~\eqref{eq:p_s}. These expressions agree with the time-domain solutions for few-photon transport obtained in~\cite{shen2007strongly,yudson2008multiphoton,shi2009lehmann,shi2011two}.

\section{\label{sec_3} Results}


\subsection{Splitting Unentangled Photons}


We first analyse the splitting efficiency for an input of two unentangled photons. In this case we can write the wavefunction as $\xi(t_{1}, t_{2}) = \sqrt{2} ~\xi(t_{1})\xi(t_{2})$. Here $\xi(t)$ is a normalized single photon wavepacket and the factor of $\sqrt{2}$ ensures that the input state is normalized under time-ordering $t_2 \geq t_1$ \cite{heuck2020photon}. We analyze two pulse profiles for the single photon input wavepackets.  The first is an exponential pulse profile such that $\xi(t)=\sqrt{2\kappa}e^{-\kappa t}$, and the second is a gaussian pulse profile given by $\xi(t) = \left(\sqrt{\frac{2}{\pi}} \kappa\right)^{\frac{1}{2}}  e^{-\kappa^{2} t^{2}}$. In both cases $\kappa$ parametrizes the bandwidth of the pulse.

\begin{figure}[t]
    \centering
    \includegraphics[width = \columnwidth]{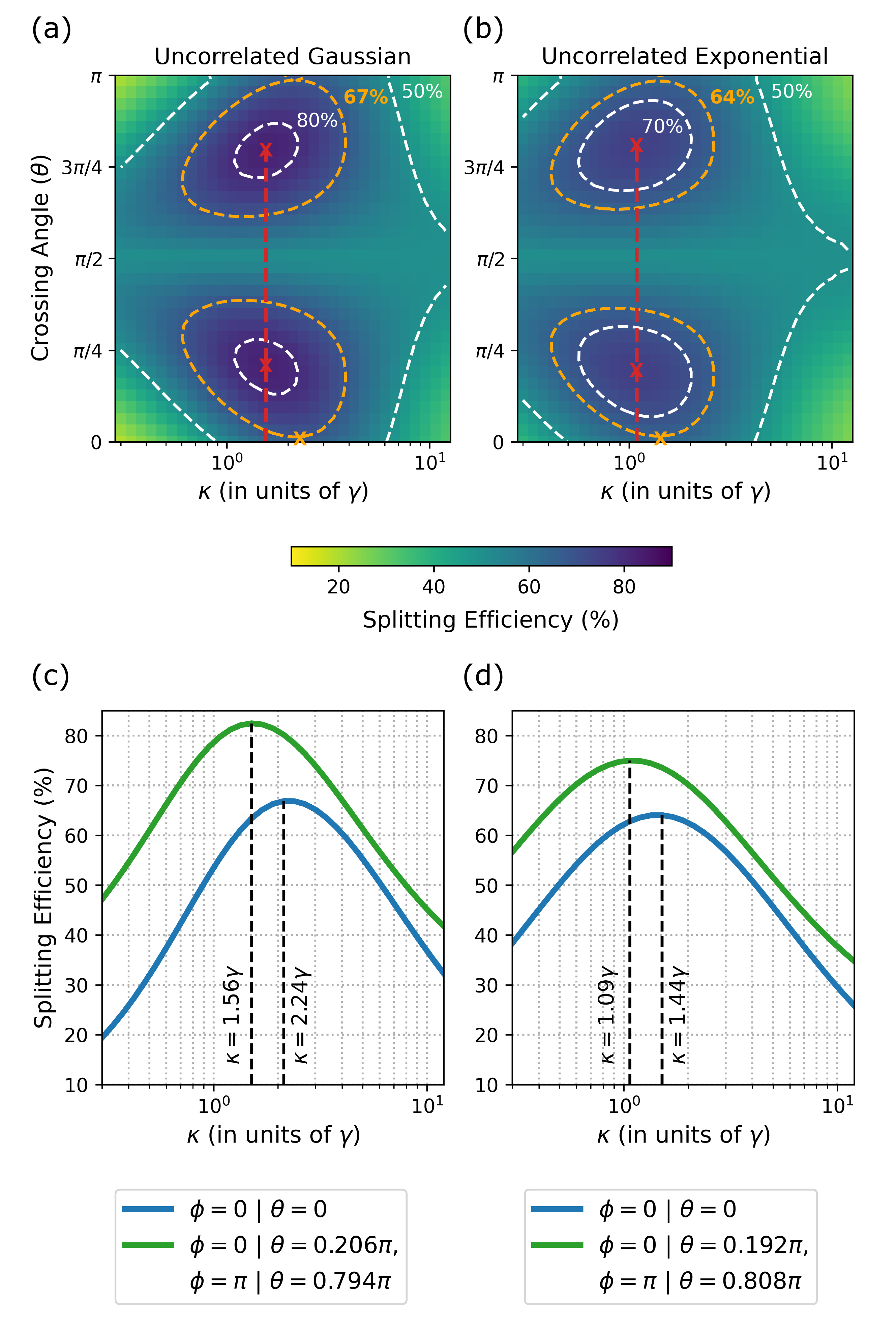}
    \caption{ \textbf{Splitting efficiency for unentangled input photons.}~\textbf{(a)} Splitting efficiency (indicated in color) for a two-photon input with unentangled Gaussian pulses, as a function of the photon bandwidth $\kappa$ and the crossing angles $\theta$. The maximum splitting efficiency is indicated in red color (82.5\%), with the orange colored contours corresponding to the blockade limited splitting efficiency (67\%). \textbf{(b)} Splitting efficiency for unentangled exponential input pulses, with the same labels as panel (a). The maximum efficiency and blockade limited efficiency are 75\% and 64\%. \textbf{(c)} Splitting efficiency for unentangled Gaussian pulses plotted as a function of photon bandwidth $\kappa$ with and without enhancement by the unitary, plotted in green and blue respectively. The peaks of these curves correspond to the maximum splitting efficiency and the blockade limited efficiency seen in panel (a). \textbf{(d)} Splitting efficiency for unentangled exponential pulses, with the same labels as panel (b). The peaks of these curves also correspond to the maximum splitting efficiency and blockade limited efficiency seen in panel (b).}
    \label{fig:unentangled}
\end{figure}




We calculate the splitting efficiency using the results of the previous section. In Appendix~\ref{sec:appendix_anasol}, we perform the full calculation for the exponential wavepacket, which leads to an analytical solution. For the Gaussian pulse it is not possible to attain an analytical expression so we  numerically calculate the splitting efficiency.  Figures~\ref{fig:unentangled}(a) and~\ref{fig:unentangled}(b) plot the resulting splitting efficiency as a function $\kappa$ and $\theta$ for uncorrelated inputs with  Gaussian and Exponential pulse profiles respectively. Since the splitting efficiency has a periodicity in $\theta$ of $\pi$, we analyse and plot only one period. For each point on the plot, we optimize the value of the interferometer input phase $\phi$ (see fig.~\ref{fig:schematic}) to obtain the maximum splitting efficiency for the values of $\theta$ and $\kappa$ corresponding to that point. We find that $\phi=0$ optimizes the splitting efficiency for all points with $\theta \leq \pi/2$ and $\phi= \pi$ optimizes the splitting efficiency for $\theta>\pi/2.$ This is true for both exponential and Gaussian pulses. Furthermore, the splitting efficiency optimized for $\phi$ is mirrored across the line $\theta=\pi/2$, such that the value at $\theta$ is the same as the value at $\pi-\theta$, where $\theta < \pi/2.$ 

The red dot denotes the maximum splitting efficiency in both plots, which is obtained at $(\theta,\phi)$ values of $(0.206\pi,0)$ for the Gaussian wavepacket and $(0.192\pi,0)$ for the exponential wavepacket. As noted in the last paragraph, there are values of $\theta>\pi/2$ which result in the same optima when the phase difference $\phi$ between the two input arms of the interferometer is set to $\pi.$ The orange contours represent the bare atom splitting efficiency calculated in ref.~\cite{rosenblum2011photon} for an Gaussian wavepacket (67\%) and exponential wavepacket (64\%). One can see that the red dot in both cases is within these orange contours and therefore achieves a higher splitting efficiency.

We next compare the optimal splitting efficiency obtained with the Mach-Zehnder interferometer to that of the bare atom. We can extract the bare atom splitting efficiency from the $\theta=0$ cross-section of the plots in figures ~\ref{fig:unentangled}(a) and~\ref{fig:unentangled}(b). In this case the unitary implements the identity transformation and we therefore recover the bare atom response.  Figs.~\ref{fig:unentangled}(c)  and~\ref{fig:unentangled}(d) plot the splitting efficiency as a function of the input pulse bandwidth for these two unitary transformations. The blue curves correspond to having no unitary on the outputs of the atom, and give the blockade limited efficiency. The green curves corresponds to the unitary transformation that optimizes the splitting efficiency. For the exponential pulse, the blockade limited efficiency is $65\%$ and occurs at $\kappa=1.44\gamma$. In contrast, the optimal splitting efficiency with the unitary is 75\%. The bandwidth $\kappa$ which achieves this global maximum is $1.09\gamma$, is therefore smaller than the optimal bandwidth which achieves the blockade limited efficiency. For the Gaussian pulse, optimal bandwidth $\kappa$  is $1.57\gamma$, which achieves an optimal splitting efficiency of 82\%. This efficiency is larger than the blockade limited efficiency of 67\%. We achieve this optimal at a smaller than that realized by the bare atom, which is $2.24\gamma$ for  a Gaussian input pulse.  Therefore, the unitary transformation fundamentally changes the time-bandwidth trade-off required to achieve optimal splitting efficiency.

\subsection{Splitting Time-Energy Entangled Photons}

We now analyse the response of our system for inputs which are time-energy entangled. These inputs have time-energy uncertainty relations which are fundamentally different from the uncorrelated inputs. Therefore, their interaction with the two level atom is also different. One way to write an time-energy entangled photon state is as follows:
\begin{equation}
\ket{\psi} = \int \d\omega_{0}~G(\omega_0)  \int \d\omega~F(\omega) \ket{\omega,\omega_{0} - \omega}
\label{eq:ent_freq}
\end{equation}
where $G(\omega_0)$ and $F(\omega)$ are general wavefunctions constrained only by the requirement for the overall normalization of the state. The above wavefunction can be expressed in the time domain as:
\begin{equation}
\ket{\psi} = \int \d t_{1}~g(t_{1})  \int \d t_{2}~f(t_{2} - t_{1}) a^{\dagger}(t_{1}) a^{\dagger}(t_{2}) \ket{0}    
\label{eq:ent_time}
\end{equation}
where $g(t_{1})$ and $f(t_{2}-t_{1})$ are Fourier transforms $G(\omega_{0})$ and $F(\omega)$. In the limit where $G(\omega_0) = \delta(\omega_0)$, when then have $g(t_{1}) = \frac{1}{2\pi}$ which achieves a perfect temporally correlated entangled state which depends only on the time difference $t_{2} - t_{1}$. We refer to such states as stationary, because the correlations only depend on the arrival time difference, and do not depend on the values of the individual time variables. A more general entangled state can introduce non-stationary behavior where correlations are time-dependent, with the dependence quantified by the function $g(t_1)$.

We first consider the specific case where $g(t_{1}) = \sqrt{2\kappa}e^{-\kappa t_{1}}$ and $f(t_{2} - t_{1}) = \sqrt{2\delta}e^{-\delta(t_{2} - t_{1})}$. Here the two wavefunctions are exponential where $\kappa$ and $\delta$ represent the bandwidths of the respective distributions. We attain the stationary limit when $\kappa\to0$. From this state, we obtain the following normalized input wavefunction:
\begin{equation}
    \xi(t_1,t_2)= 2 \sqrt{\kappa  \delta }~e^{-\kappa  t_1} e^ {-\delta  (t_2 -t_1)}
    \label{eq:expent}
\end{equation}
where $\xi(t_1,t_2)$ is defined in eqn.~\eqref{eq:init_state}. We begin with the above input wavefunction because it leads to an analytical solution. We give this full analytical solution in Appendix~\ref{sec:appendix_anasol}. We are only interested in the stationary limit, which we obtain by taking $\kappa \to \infty$. In this limit, the expression for splitting efficiency is identical to the one obtained for maximally entangled states generated with a three level atomic cascade \cite{rosenblum2011photon}. Maximizing this expression with respect to the bandwidth $\delta$ with and without the linear optical unitary yields splitting efficiencies of $90\%$ and $77\%$ respectively. 

\begin{figure}[t!]
    \centering
    \includegraphics[width = \columnwidth]{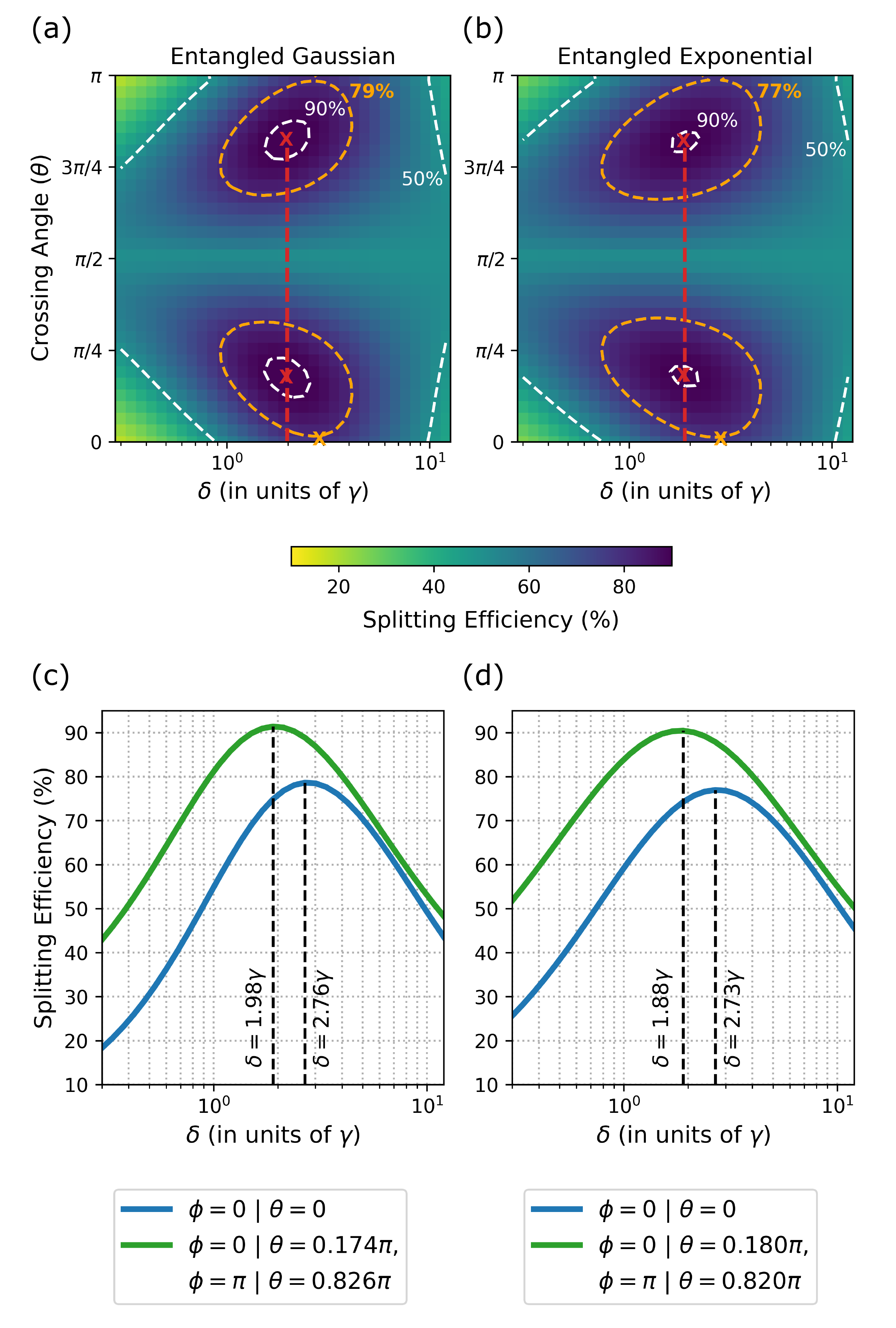}
    \caption{ \textbf{Splitting efficiency for entangled input photons.}~\textbf{(a)} Splitting efficiency (indicated in color) for a two-photon input with entangled Gaussian pulses, as a function of the photon bandwidth $\kappa$ and the crossing angles $\theta$. The maximum splitting efficiency is indicated in red color (91.5\%), with the orange colored contours corresponding to the blockade limited splitting efficiency (79\%). \textbf{(b)} Splitting efficiency for entangled exponential input pulses, with the same labels as panel (a). The maximum efficiency and blockade limited efficiency are 90\% and 77\%. \textbf{(c)} Splitting efficiency for entangled Gaussian pulses plotted as a function of photon bandwidth $\kappa$ with and without enhancement by the unitary, plotted in green and blue respectively. The peaks of these curves correspond to the maximum splitting efficiency and the blockade limited efficiency seen in panel (a). \textbf{(d)} Splitting efficiency for entangled exponential pulses, with the same labels as panel (b). The peaks of these curves also correspond to the maximum splitting efficiency and blockade limited efficiency seen in panel (b). }
    \label{fig:entangled}
\end{figure}



Our analysis for the input state given by eqn.~\eqref{eq:expent}  suggests that a stationary time-energy correlated input $\xi_{s}(|t_2-t_1|)$ can significantly improve the splitting efficiency over uncorrelated inputs. Stationary inputs with different pulse profiles could yield further improvements.  In the previous section, uncorrelated inputs with a Gaussian pulse profile yielded a bigger maximum for splitting efficiency than Exponential pulses. We therefore consider the following input state:
\begin{equation}
  \ket{\Psi_{\mathrm{in}}}=\int_{-L}^{L} \d t_{1}\int_{t_{1}}^{\infty} \d t_{2}  \sqrt{\frac{\delta}{\pi L}} e^ {-\frac{\delta}{2}  (t_2 -t_1)^2} \hat{a}^\dagger_{\mathrm{in}}(t_{1})\hat{a}^\dagger_{\mathrm{in}}(t_{2})\ket{0}  
\label{eq:entGauss}
\end{equation}

Note that this state corresponds to substituting a Gaussian $F(\omega_1)=e^{-2\omega_1^2/\delta^2}$ in eq. 12. Here, $\delta\geq0$ gives the bandwidth of the Gaussian and hence, of the input pulse. We note that in this case the stationary limit corresponds to $L\to \infty.$ For this input state, the splitting efficiency is calculated via numerical integration. 

Figures 3(a) and 3(b) plot the resulting splitting efficiency as a function $\delta$ and $\theta$ for stationary entangled inputs with  Gaussian and Exponential pulse profiles respectively.  Since the splitting efficiency is periodic in $\theta$, we plot only one period. 
Each point on these plots corresponds to optimizing the splitting efficiency with respect to $\phi$. We obtain the same dependence on $\phi$ as for the uncorrelated inputs such the splitting efficiency optimized for $\phi$ is mirrored across $\theta=\pi/2.$ The red dots correspond to the splitting efficiencies obtained by optimizing the linear optical unitary, which are $91.5\%$ and $90\%$ for entangled Gaussian and exponential pulses respectively. These lie within the orange contours that represent the blockade limited splitting efficiencies. We note that the blockade limited splitting efficiency of $77\%$ corresponds to the value obtained in ref. \cite{rosenblum2011photon} for input photons generated by a three level atomic cascade.

We now compare the splitting efficiency with obtained with the bare atom and the optimal linear optical unitary for both exponential and Gaussian entangled pulses. To make this comparison, we plot these two cases for Gaussian and exponential pulses in figures 3(c) and 3(d) respectively. The blue curves correspond to having no unitary on the outputs of the atom, and give the blockade limited efficiency. The green curves corresponds to the unitary transformations which optimize the splitting efficiency. For the entangled exponential, the blockade limited efficiency of $77\%$ occurs at the bandwidth $\delta=2.73\gamma$.  The bandwidth $\delta$ needs to be reduced to $1.88\gamma$ to obtain the maximum splitting efficiency with the optimized linear optical unitary. For the Gaussian pulse, we also observe a reduction in the optimal bandwidth of the input pulse in going from the bare atom to adding the optimized linear optical unitary transformation after the atom. The optimal bandwidths in the two cases are $2.76\gamma$ and $1.98\gamma$ respectively. Therefore, for entangled exponential and Gaussian inputs, the linear optical unitary changes the time-bandwidth tradeoff required to optimize the splitting efficiency. This is in line with our findings for uncorrelated inputs in the previous section. 

\section{Optimal Splitting via Temporally Shaped Photons \label{sec_4}}

In the previous section we assumed that the stationary entangled photon wavefunction takes on the specific form of a Gaussian or exponential.  In this section we employ optimization to obtain an optimal pulse shape that achieves the globally optimum splitting efficiency.  This optimal waveform represents a strong upper limit for the splitting efficiency.

\begin{figure}
    \centering
    \includegraphics[width = 1.0\columnwidth]{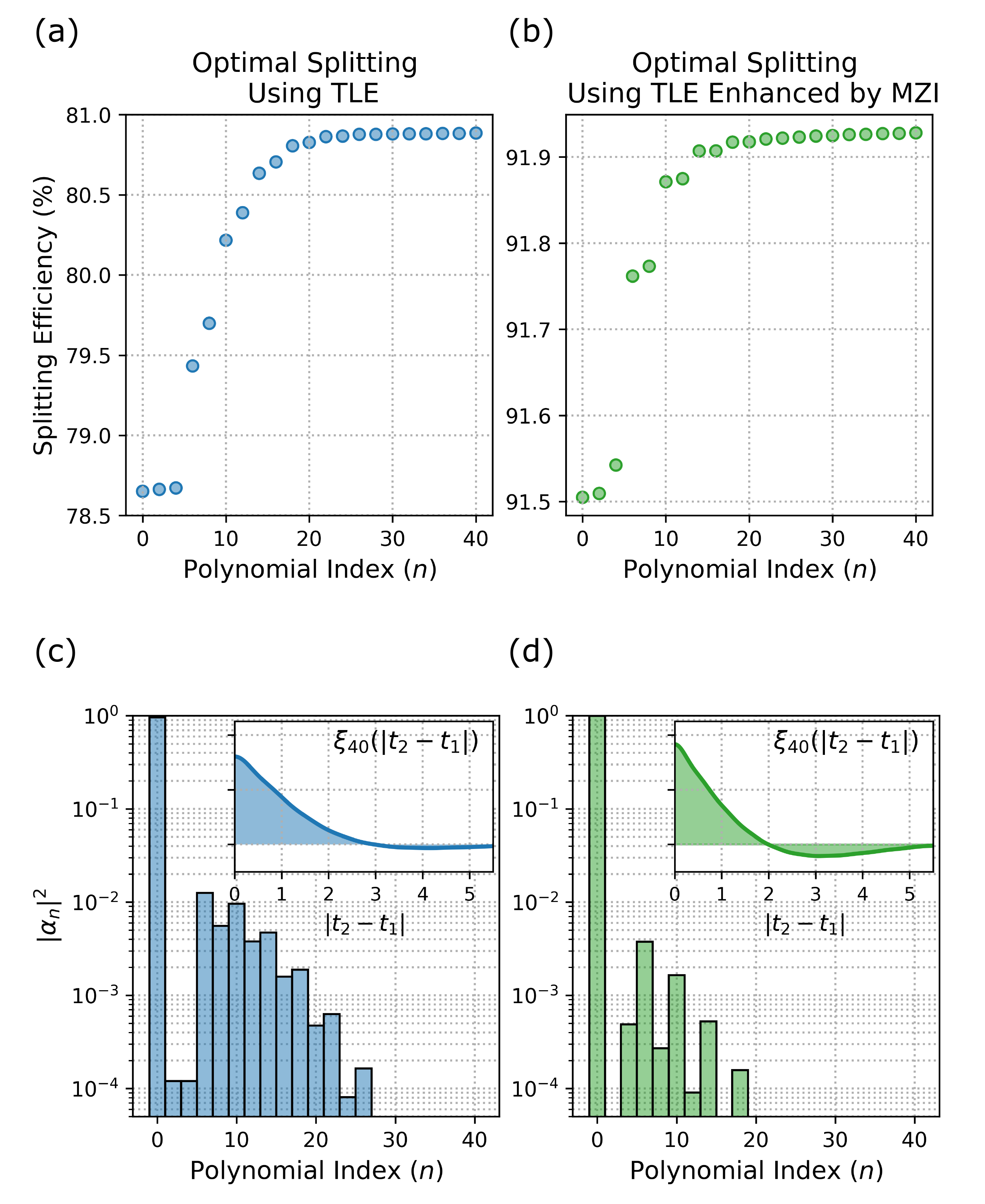}
    \caption{\textbf{Optimal splitting efficiency with temporally shaped input pulses.}~\textbf{(a)} Splitting efficiency of the bare two-level emitter as a function of the number of Gauss-Hermite polynomials contributing to the optimal pulse profile. \textbf{(b)} Splitting efficiency of the of the emitter enhanced by the interferometer as a function of the number of Gauss-Hermite polynomials contributing to the optimal pulse profile.
    \textbf{(c)} Contributions of each element to the optimal pulse profile to maximize the splitting efficiency of the bare two-level emitter. The absolute value squared of the coefficients $\alpha_{n}$ is plotted in the bar plot, with the first polynomial (Gaussian component) contributing the maximum amount. The inset shows the optimal pulse profile. \textbf{(d)} Contributions of each element to the optimal pulse profile to maximize the splitting efficiency of the emitter enhanced by the interferometer. The labels are the same as in panel (c), where the first polynomial (Gaussian component) contributes the maximum amount. The inset shows the optimal pulse profile.}
    \label{fig:optimal}
\end{figure}


To optimize the entangled photon pulse shape, we expand the stationary wavefunction $\xi_s(|t_2-t_1|)$ in a complete expansion basis.  Typically, such pulse shaping can be achieved by modifying a finite number of Fourier components of pulses~\cite{weiner2011ultrafast}. Instead of the Fourier basis, we choose to expand the pulse profile in the Gauss-Hermite basis, given by: 
\begin{equation}
    \xi_{s}(\tau)=\sum_{n=0}^{N/2} \alpha_{n} \mathrm{H}_{2n}(\tau),  \hspace{0.5cm}   \text{for even $N$}
    \label{eq:GH_expand}
\end{equation}
where $\mathrm{H}_n(x)$ are the normalized Gauss-Hermite polynomials such that $\sum_{n = 0}^{N/2}|\alpha_{n}|^{2} = 1$ and $N/2$ gives the number of terms in the basis expansion. The terms $\alpha_{n}$ are the coefficients of the respective polynomials, ensuring that the function $\xi_{s}(\tau)$ is normalized. We chose the Gauss-Hermite basis of polynomials because the first term in this set is purely Gaussian, which has been shown to give us a high splitting efficiency of 91.5\%. We therefore expect that higher order terms will add only small corrections, and we will only need to keep a few of them to come close to the global limit. Note that we keep only even terms in the sum in eqn.~\eqref{eq:GH_expand}, which corresponds to expanding $\xi_s(\tau)$ over only the even Hermite-Gauss functions. We can ignore the odd Hermite-Gauss functions without any loss of generality because $\tau \geq 0$, so we can express any function on the positive time axis using only even functions of time, up to $L^2$ convergence.

Since the scattering amplitudes $\psi_\mathrm{pq}$ and $\psi_\mathrm{lm}$ from eqn.~\eqref{eq:psi_pq} and eqn.~\eqref{eq:psi_lm} are linear in the input pulse, the output two photon wavefunction for an arbitrary input pulse $\xi(\tau')$ denoted by $\psi_\mathrm{pq}\left( \xi \right) $ can be written in as:

\begin{equation}
    \psi_\mathrm{pq} \left( \xi \right) = \sum_{n = 0}^{N/2} \alpha_{n}\psi_\mathrm{pq}\left( \mathrm{H}_{2n} \right)
\end{equation}

The splitting efficiency is obtained by integrating $|\psi_\mathrm{cd}|^2 + |\psi_\mathrm{dc}|^2$ over the output photon creation times $\tau_{1}$ and $\tau_{2}$. For example, $\Gamma^\mathrm{cd}\left( \xi \right) = |\psi_\mathrm{cd}\left( \xi \right)|^2$ is given by:

\begin{equation}
\Gamma^{\mathrm{cd}} \left( \xi \right)= \sum_{m=0}^{N/2}\sum_{n=0}^{N/2} \alpha_{m}^{*}\alpha_{n}\psi_\mathrm{cd}\left( \mathrm{H}_{2m} \right) \psi_\mathrm{dc}\left( \mathrm{H}_{2n} \right) 
\label{eq:2p_basis_def}
\end{equation}
where we use the fact that the scattering amplitudes for real valued input states are real (see eqns.~\eqref{eq:scat_amps}). This can be rewritten as:

\begin{equation}
\Gamma^{\mathrm{cd}}\left( \xi \right)  = \vec{\alpha}^\dagger \boldsymbol{\rho}^{\mathrm{cd}} \left( \mathrm{H}_{2n} \right) \vec{\alpha}
\end{equation}
where $\vec{\alpha}$ is a column vector with entries $\alpha_1$ through $\alpha_{N/2}$ and $\boldsymbol{\rho}^{\mathrm{cd}} \left( \mathrm{H}_{2n} \right)$ is the probability density matrix where the entries are determined by the choice of basis, i.e., $\left( \mathrm{H}_{2n} \right)$ from~\eqref{eq:2p_basis_def} . We note that $\boldsymbol{\rho}^{\mathrm{cd}}$ is a symmetric matrix because $\psi_\mathrm{cd}\left( \mathrm{H}_{2m}\right) \psi_\mathrm{dc}\left( \mathrm{H}_{2n} \right)$ is commutative as can be seen from~\eqref{eq:2p_basis_def}. A similar argument can be made for $\rho^{\mathrm{dc}}$ and therefore the probability density $\boldsymbol{\rho}_\mathrm{s} \left( \xi \right) = \Gamma^{\mathrm{cd}} \left( \xi \right) +\Gamma^{\mathrm{dc}} \left( \xi \right) $ of the splitting the photons to output ports $c$ and $d$ takes a similar form:

\begin{equation}
\boldsymbol{\rho}_{\mathrm{s}}\left( \xi \right) = \vec{\alpha}^{\dagger} \boldsymbol{\rho}_{\mathrm{s}} \left( \mathrm{H}_{2n} \right) \vec{\alpha}
\label{eq:prob_density_xi}
\end{equation}
Integrating $\boldsymbol{\rho}_{\mathrm{s}} \left(\xi \right)$ over $\tau_1$ and $\tau_2$ is equivalent to integrating~\eqref{eq:prob_density_xi} element-wise over $\tau_1$ and $\tau_2$, which yields another symmetric matrix that we denote as $\boldsymbol{R}$. Note that the diagonal entries of $\boldsymbol{R}$ are simply the splitting efficiencies for the different basis elements. Therefore the splitting efficiency for the input state $\xi(\tau)$ can be written as:

\begin{equation}
 P_{\mathrm{S}}(\vec{\alpha})=\vec{\alpha}^\dagger \boldsymbol{R} \vec{\alpha}
\end{equation}

Note that $\textbf{R}$ is a symmetric positive semi-definite matrix because $P_{\mathrm{S}}(\vec{\alpha}) \geq 0$ by definition. 

We claim that the maximum $P_{\mathrm{S}}(\vec{\alpha})$ is obtained for $\vec{\alpha}_{\mathrm{max}}$ which is the eigenvector of $\boldsymbol{R}$ corresponding to its maximum eigenvalue. To see this we first digonalize $\boldsymbol{R} = \boldsymbol{U}^{\mathrm{T}}\boldsymbol{D}\boldsymbol{U}$ using the spectral theorem for symmetric matrices. Since, $\boldsymbol{R}$ is positive semi-definite, all entries of the the diagonal matrix $\boldsymbol{D}$ are positive. The splitting efficiency can be rewritten as:

\begin{equation}
P_\mathrm{S}(\vec{\beta})=\vec{\beta}^\dagger \boldsymbol{D} \vec{\beta}
\end{equation}

where $\vec{\beta} = \boldsymbol{U} \vec{\alpha}$ and the normalization constraint $\vec{\alpha}^\dagger\vec{\alpha}$ is equivalent to the constraint $\vec{\beta}^\dagger \vec{\beta}$. Therefore, $P_S(\vec{\beta})$ is clearly maximum when the vector $\vec{\beta}_{\mathrm{max}}$ has zero valued entries everywhere except the position corresponding to the maximum diagonal entry of $\boldsymbol{D}$, which is the maximum eigenvalue of $\boldsymbol{R}$. Therefore, the vector $\vec{\alpha}_{\mathrm{max}}=\boldsymbol{U} \vec{\beta}_{\mathrm{max}}$ is the eigenvector of $\boldsymbol{R}$ with the maximum eigenvalue. 

To find the maximum splitting efficiency, we construct the matrix $\boldsymbol{R}$ for different number of terms in the basis expansion. The maximum eigenvalue of $\boldsymbol{R}$ then gives the maximum splitting efficiency.  Using the corresponding eigenvector $\vec{\alpha}_{\mathrm{max}}$ of $\boldsymbol{R}$ in (eqn.~\eqref{eq:GH_expand}) gives the optimal pulse shape in each case.

Figs.~\ref{fig:optimal}(a) and~\ref{fig:optimal}(b) show the dependence of the maximum splitting efficiency on the number of Gauss-Hermite polynomial in the basis expansion of the input wavefunction. Fig.~\ref{fig:optimal}(a) plots this for the case where the bare two level atom is used to split the input photons. The maximum splitting efficiency increases monotonically from 78.5\% and saturates at $\sim$ 81\%. Fig.~\ref{fig:optimal}(b) shows the case where the optimal linear optical unitary enhances the maximum splitting efficiency from 91.5\% to $\sim 92\%.$ In both cases, when the input pulse is Gaussian (i.e., when $n = 0$), the splitting efficiency is very close to the value it ultimately saturates to as we add more terms to the basis expansion of the input wavefunction. This suggests that the Gaussian pulse shape is very close to the optimal pulse shape of the input wavefunction. 

Figs.~\ref{fig:optimal}(c) and~\ref{fig:optimal}(d) illustrate the contributions of each basis element to the optimal pulse profile. Fig.~\ref{fig:optimal}(c) plots the index of the polynomial against its corresponding weight squared ($|\alpha_{n}|^{2}$) for splitting with the bare two-level emitter. Fig.~\ref{fig:optimal}(d) plots the same parameters for the case when the splitting is enhanced by the Mach-Zehnder Interferometer. The insets in each of these figures plots the optimal pulse with a total of 40 basis elements contributing to its profile. We see that the first polynomial ($n = 0$, which is purely Gaussian) contributes the largest weight to the optimal pulse in both cases, indicating that the Gaussian comes very close to the optimal pulse profile. In the case of the bare emitter, $|\alpha_{0}|^{2} \sim 0.958$, and in the case of splitting enhanced by the interferometer, $|\alpha_{0}|^{2} \sim 0.992$. 
We truncate the number of basis elements to 40 because we find that the weights of the higher order basis elements drops significantly and does not contribute to the profile of the optimal pulse. This is also indicated by the plateauing of the splitting efficiencies in figs.~\ref{fig:optimal}(a) and (b). 

In this section, our calculations were based expanding the pulse profile in the Gauss-Hermite basis. This is motivated by our results from section~\ref{sec_3}, where the first element of the basis (a pure Gaussian) attained a high splitting efficiency. In general, the choice of basis can be arbitrary as long as it is  complete over stationary wavefunctions and is orthonormal. In an alternate basis, the splitting efficiency matrix $\textbf{R}'$ will be determined by the elements of that basis. $\textbf{R}'$ is still a symmetric positive semidefinite matrix in the new basis. A change of basis implies that $\textbf{R}'$ can be related to $\textbf{R}$ through a unitary transformation. The same procedure of finding the maximum eigenvalue presented above can be followed in the new basis as well. We expect that any complete basis would converge to the same solution presented here.

\section{\label{sec_5}Discussion and Conclusion}



In conclusion, we have presented an extensive theoretical analysis of the splitting of two indistinguishable photons to spatially distinct output channels using a two-level emitter followed by a Mach-Zehnder interferometer. Through optimization of the input pulse shape and the phases of the interferometer, we obtain a splitting efficiency of $92\%.$ This is a substantial improvement over the optimal splitting efficiency with just the two-level emitter, which is close to $81\%.$ Our results exceed the maximum splitting efficiency of $77\%$ with a two-level emitter calculated in ref. \cite{rosenblum2011photon}. Our results also suggest that Gaussian pulse profiles extract close to optimal nonlinear response from the two level atom. This supports our intuition because Gaussian pulses have the minimal time-bandwidth product. 
  

One limitation of our analysis is that it was performed in the weak coupling regime of cavity quantum electrodynamics, where the cavity modes could be adiabtically eliminated. However, in the strong coupling regime, the cavity modes cannot be adiabatically eliminated and the level structure of the atom-cavity system is given by the Jaynes-Cummings ladder. Nevertheless, we expect to obtain the same optimal splitting efficiencies in the strong coupling regime as were obtained in our current work. This hypothesis is supported by the simulations performed in ref.~\cite{rosenblum2011photon}. There the authors showed that the optimal splitting efficiency in the strong coupling regime occurs when the coupling rate $g$ between the atom and the cavity is sufficiently large such that the interaction between the incoming photons and the atom-cavity system is dictated by the ground state and the first excited state of the Jaynes-Cummings ladder. Therefore, we expect to recover the same results as in our current analysis, but with the atomic bandwidth $\gamma$ replaced by bandwidth of the first excited state of the Jaynes-Cummings ladder i.e. $\kappa+\gamma$, where $\kappa$ is the bandwidth of the cavity.

This work has proved that the use of optimal interference schemes can scale past the limit on splitting efficiency imposed by the photon blockade. From a hardware standpoint, the proposed model utilizes only a single two-level emitter, weakly coupled to a waveguide and a Mach-Zehnder Interferomter, which are both passive and can be fabricated easily with well-established lithography techniques. While the nonlinear effect of the emitter and interferometer is still not perfect, near term improvements in control schemes, such as dynamic coupling~\cite{heuck2020photon, krastanov2022controlled} and dispersion engineering~\cite{notaros2017programmable}, would enable high-fidelity quantum operations. Alternatively, active control schemes such as introducing a time-varying linear optical unitary or non-markovian coupling to waveguide or environment modes may also expected to increase the splitting efficiency. 

Networks of linear optical unitaries with nonlinear interactions have been a growing area of interest due to their versatility. In particular, cascading multiple two-level emitters~\cite{yang2022deterministic} has been shown to realize perfect photon sorting. Such networks combined with beam-splitter meshes have also been studied with idealized single-mode Kerr interactions instead of a two-level atom as the nonlinearity \cite{steinbrecher2019quantum, ewaniuk2023imperfect} for state generation, quantum repeater nodes and bell-state analysis.  Optimization of such networks may improve the fidelities of operations besides photon-photon splitting that we considered here. Our results highlight the potential to generate improved quantum optical interaction by combining single-photon nonlinearities with linear optical systems, with potential applications in photon quantum information processing and quantum simulation.

\subsection*{Materials and Correspondence}
All requests for code and data should be made to H.S. at \url{hsingh16@umd.edu}. 

\subsection*{Competing Interests}
The authors declare no competing interests.

\begin{acknowledgments}
The authors would like to acknowledge financial support from the National Science Foundation (grants \#OMA1936314, \#OMA2120757, \#PHYS1915375 and \#ECCS1933546), the AFOSR grant \#FA23862014072, and the Maryland-ARL Quantum Partnership (W911NF1920181).
\end{acknowledgments}

\bibliography{main}

\clearpage

\begin{widetext}
\section*{Appendix}
\end{widetext}
\appendix

\section{Derivation of Scattering Amplitudes}
\subsection{Theoretical Foundations to Calculate the Scattering Amplitudes}
We begin by introducing the formalism and formulae required to derive the scattering amplitudes. In order to do so, we note:
\begin{align}
    \psi_\mathrm{pq}(\tau_{1},\tau_{2}) & = \bra{0}\hat{p}_{\mathrm{out}}(\tau_{2})\hat{q}_{\mathrm{out}}(\tau_{1})\ket{\psi_{0}} \nonumber \\ & =  \int_{-\infty}^{\infty} \d t_{1}\int_{t_{1}}^{\infty} \d t_{2}~\xi(t_{1},t_{2}) G_\mathrm{pq}(\tau_1,\tau_2; t_1, t_2)   
    \label{eq:appendix_psi_pq}
\end{align}
where $\{ \hat{p}, \hat{q} \} \in \{ \hat{c}, \hat{d} \}$ are the output modes of the bare two-level emitter, as defined in the main text, and $G_\mathrm{pq}(\tau_1,\tau_2;t_1,t_2)$ is defined by:
\begin{equation}
   G_\mathrm{pq}(\tau_{1},\tau_{2};t_{1},t_{2})= \bra{0}\hat{p}_{\mathrm{out}}(\tau_{2})\hat{q}_{\mathrm{out}}(\tau_{1})\hat{a}^\dagger_{\mathrm{in}}(t_{1})\hat{a}^\dagger_{\mathrm{in}}(t_{2})\ket{0}
\end{equation}

Therefore, to calculate the desired scattering amplitudes $\psi_\mathrm{pq}$, we first calculate $G_\mathrm{pq}$ and then use eqn.~\eqref{eq:appendix_psi_pq}.In order to calculate $G_\mathrm{pq}$, we first recall the result from ref.~\cite{xu2015input} that will be used repeatedly in this section. Throughout this section the time orderings $t_2 \geq t_1$ and $\tau_2 \geq \tau_1$ are assumed. 

\begin{subequations}
\begin{equation}
    \bra{0}\hat{\sigma}(\tau_{1})\hat{\sigma}^{\dagger} (t_{1})\ket{0} = \bra{0}\Tilde{\sigma}(\tau_{1})\Tilde{\sigma}^{\dagger}(t_{1})\ket{0}  
\end{equation}
\begin{multline}
\bra{0}\hat{\sigma}(\tau_{1})\hat{\sigma}(\tau_{2})\hat{\sigma}^{\dagger} (t_{2})\hat{\sigma}^{\dagger} (t_{1})\ket{0} = \\
     \bra{0}\Tilde{\sigma}(\tau_{1})\Tilde{\sigma}(\tau_{2})\Tilde{\sigma}^{\dagger}(t_{2})\Tilde{\sigma}^{\dagger}(t_{1})\ket{0}  
\end{multline}
\label{eq:appendix_sigma_relations}
\end{subequations}
with
\begin{equation}
\Tilde{\sigma}(t)= e^{i\hat{H}_\mathrm{eff}t}\hat{\sigma} e^{-i\hat{H}_\mathrm{eff}t}   
\label{eq:sigma_tilde}
\end{equation}
where
\begin{equation}
    \hat{H}_\mathrm{eff}=-2i\gamma\hat{\sigma}^\dagger\hat{\sigma}
    \label{eq:eff_Hamil}
\end{equation}

Using Eqs.~\eqref{eq:sigma_tilde} and~\eqref{eq:eff_Hamil} and the properties of the operators $\hat{\sigma}$ and $\hat{\sigma}^\dagger$, we can calculate the amplitudes corresponding to eqns.~\eqref{eq:appendix_sigma_relations}. In eqn.~\eqref{eq:appendix_sigma_relations}(a), $\tau_{1} \geq t_{1}$ is the only possible time ordering because the atom must be raised to the excited state before it is lowered to the ground state. In eqn.~\eqref{eq:appendix_sigma_relations}(a), $\tau_{2} \geq t_{2}\geq \tau_{1}\geq t_{1}$ is the only possible ordering, because before being raised by $\hat{\sigma}^\dagger(t_2)$, the atom must be lowered. Therefore, we have:

\begin{subequations}
\begin{equation}
    \bra{0}\hat{\sigma}(\tau_{1})\hat{\sigma}^{\dagger}(t_{1})\ket{0} = e^{-2\gamma(\tau_{1}-t_{1})} \Theta(\tau_{1}-t_{1})
\end{equation}
and
\begin{multline}
    \bra{0}\hat{\sigma}(\tau_{1})\hat{\sigma}(\tau_{2})\hat{\sigma}^\dagger(t_{2})\hat{\sigma}^{\dagger}(t_{1})\ket{0} = \\ 
    e^{-2\gamma(\tau_{2}-t_{2})}  e^{-2\gamma(\tau_{1}-t_{1})} \\ \times \Theta(\tau_{2}-t_{2}) \Theta(\tau_{2}-t_{1})  \Theta(\tau_{1}-t_{1}) 
\end{multline}
\label{eq:appendix_heaviside_def}
\end{subequations}
where $\Theta$ is the Heaviside step function and ensures the time ordering. Before we proceed to the calculations of $G_\mathrm{pq}$, we also recall the quantum causality conditions :

\begin{subequations}
\begin{equation}
    \left[ \hat{\sigma}(t), \hat{I}(\tau) \right] = \left[ \hat{\sigma}^\dagger(t), \hat{I}(\tau)\right] = 0, \mathrm{for}\,\,t\leq \tau 
\end{equation}
\begin{equation}
     \left[ \hat{\sigma}(t), \hat{O}(\tau)\right]=\left[ \hat{\sigma}^\dagger(t), \hat{O}(\tau) \right]=0, \mathrm{for}\,\, t\geq \tau 
\end{equation}
\label{eq:appendix_causality}
\end{subequations}
where $\hat{I}$ stands for \textit{input} annihilation or raising operators, and $\hat{O}$ stands for \textit{output} annihilation or raising operators.

\subsection{Exact Solutions for the Scattering Amplitudes}

In this section, we derive the exact analytical form of the scattering amplitudes presented in the main text of this manuscript in eqn.~\eqref{eq:scat_amps}. We first calculate the scattering amplitudes for two reflection events $G_{\mathrm{bb}}$, which gives: 

\begin{multline}
\bra{0}\hat{b}_{\mathrm{out}}(\tau_{2})\hat{b}_{\mathrm{out}}(\tau_1)\hat{a}^{\dagger}_{\mathrm{in}}(t_{1})\hat{a}^{\dagger}_{\mathrm{in}}(t_{2})\ket{0} = \\
2\gamma\bra{0}\hat{\sigma}(\tau_{2})\hat{\sigma}(\tau_{1})\hat{a}^{\dagger}_{\mathrm{in}}(t_{1})\hat{a}^{\dagger}_{\mathrm{in}}(t_{2})\ket{0}  
\label{eq:appendix_2p_ref}
\end{multline}

We first note here that the only time ordering which gives a non-zero result is $\tau_2 \geq t_2 \geq \tau_1\geq t_1$. This is because other possible time orderings of eqn.~\eqref{eq:appendix_2p_ref} result in 0. Therefore we have:

\begin{align}
    G_\mathrm{bb} & = 2\gamma\bra{0}\hat{\sigma}(t_2)\hat{a}^\dagger_{\mathrm{in}}(\tau_{2})\hat{\sigma}(t_1)\hat{a}^\dagger_{\mathrm{in}}(\tau_{1})\ket{0} \nonumber \\
    & = 4\gamma^2\bra{0}\hat{\sigma}(t_2)\hat{\sigma}^\dagger(\tau_{2})\hat{\sigma}(t_1)\hat{\sigma}^\dagger(\tau_{1})\ket{0}
\end{align}
\begin{align}
    \implies G_{\mathrm{bb}} & = \nonumber \\ 
    & 4\gamma^2e^{-2\gamma(t_2-\tau_2)}  e^{-2\gamma(t_1-\tau_2)} \nonumber \\ 
    & \times  \Theta(t_2-\tau_2) \Theta(\tau_2-t_1)  \Theta(t_1-\tau_1) 
    \label{eq:appendix_Grr}
\end{align}
where the first step follows from applying the causality condition~\eqref{eq:appendix_causality}(a) on the result of eqn.~\eqref{eq:appendix_2p_ref}. The second step follows from using the input-output relation $\hat{a}_\mathrm{in}^\dagger(t) = \hat{a}_\mathrm{out}^\dagger(t)+\sqrt{2\gamma}\hat{\sigma}^\dagger(t)$ and the causality condition~\eqref{eq:appendix_causality}(b) and the last step follows from eqn.~\eqref{eq:appendix_heaviside_def}(b). Using the result for $G_\mathrm{bb}$ in eqn.~\eqref{eq:appendix_psi_pq}, we obtain:
\begin{multline}
    \psi_{\mathrm{bb}}(\tau_1,\tau_2)=4\gamma^2e^{-2\gamma(\tau_1+\tau_2)} \times \\ \int_{\tau_1}^{\tau_2} \d \tau_2 \int_{-\infty}^{\tau_1} \d t_1~e^{2\gamma(t_1+t_2)} \xi(t_1,t_2)
\end{multline}

Now, we move onto calculating $G_\mathrm{ba}$. To do so, we begin by observing that:
\begin{align}
    G_{\mathrm{ba}} & = \bra{0}\hat{a}_{\mathrm{out}}(\tau_{2})\hat{b}_{\mathrm{out}}(\tau_{1})\hat{a}^\dagger_{\mathrm{in}}(t_{1})\hat{a}^\dagger_{\mathrm{in}}(t_{2})\ket{0} \nonumber \\ 
    & = -\sqrt{2\gamma}\bra{0}\hat{a}_{\mathrm{in}}(t_2)\hat{\sigma}(t_1)\hat{a}^\dagger_{\mathrm{in}}(\tau_{1})\hat{a}^\dagger_{\mathrm{in}}(\tau_{2})\ket{0} \nonumber \nonumber \\ 
    & + 2\gamma\bra{0}\hat{\sigma}(t_2)\hat{\sigma}(t_1)\hat{a}^\dagger_{\mathrm{in}}(\tau_{1})\hat{a}^\dagger_{\mathrm{in}}(\tau_{2})\ket{0}
\end{align}
where the equality follows from the input-output relations. We note that the second term on the right hand side of the equality is $G_\mathrm{bb}$. Therefore, we only need to calculate the first term. Using the causality conditions~\eqref{eq:appendix_causality}(a) and~\eqref{eq:appendix_causality}(b) and the relation  $\hat{a}_\mathrm{in}^\dagger(t) = \hat{a}_\mathrm{out}^\dagger(t) +\sqrt{2\gamma}\hat{\sigma}^\dagger(t)$, we get:
\begin{multline}
    \bra{0}\hat{a}_{\mathrm{in}}(t_2)\hat{\sigma}(t_1)\hat{a}^\dagger_{\mathrm{in}}(\tau_{1})\hat{a}^\dagger_{\mathrm{in}}(\tau_{2})\ket{0} = \\
    e^{-2\gamma(t_1-\tau_1)} \Theta(t_1-\tau_1)\delta(t_2-\tau_2) 
\end{multline}
Substituting this result into eqn.~\eqref{eq:appendix_Gtr}, we obtain $G_\mathrm{ba}$ as:
\begin{align}
    \implies G_{\mathrm{ba}} & = -2\gamma e^{-2\gamma(t_1-\tau_1)} \Theta(t_1-\tau_1)\delta(t_2-\tau_2) \nonumber \\ & + G_{\mathrm{bb}} (\tau_{1}, \tau_{2})
    \label{eq:appendix_Gtr}
\end{align}

As done above using the input-output relations and the causality conditions, the scattering amplitude $\psi_{\mathrm{ba}}$ is given by: 

\begin{align}
    \psi_{\mathrm{ba}}(\tau_1,\tau_2) & = -2\gamma e^{-2\gamma \tau_1} \int_{-\infty}^{\tau_1} \d t_1~e^{2\gamma t_1}\xi(t_1, \tau_2) \nonumber \\ & + \psi_{\mathrm{bb}} (\tau_{1}, \tau_{2})
\end{align}

Now we calculate $\mathrm{G}_\mathrm{ab}$ which corresponds to the first photon getting transmitted to the output port $\hat{a}_\mathrm{out}$, and the second photon getting reflected by the atom to $\hat{b}_\mathrm{out}.$ 

\begin{align}
    G_{\mathrm{ab}} & = \bra{0}\hat{b}_{\mathrm{out}}(\tau_{2})\hat{a}_{\mathrm{out}}(\tau_{1})\hat{a}^\dagger_{\mathrm{in}}(t_{1})\hat{a}^\dagger_{\mathrm{in}}(t_{2})\ket{0} \nonumber \\ 
    & = -\sqrt{2\gamma}\bra{0}\hat{\sigma}(\tau_2)\hat{a}_{\mathrm{in}}(\tau_1)\hat{a}^\dagger_{\mathrm{in}}(t_{1})\hat{a}^\dagger_{\mathrm{in}}(t_{2})\ket{0} \nonumber \\ 
    & + 2\gamma\bra{0}\hat{\sigma}(\tau_2)\hat{\sigma}(\tau_1)\hat{a}^\dagger_{\mathrm{in}}(t_{1})\hat{a}^\dagger_{\mathrm{in}}(t_{2})\ket{0}
    \label{eq:appendix_G_ba_1}
\end{align}
Using input-output relations and the commutation relation $[\hat{a}_{\mathrm{in}}(\tau), \hat{a}^\dagger_{\mathrm{in}}(t)]=\delta(t - \tau)$, this can be simplified to:
\begin{align}
    & G_{\mathrm{ab}} = \nonumber \\ 
    & -\sqrt{2\gamma}\bra{0}\hat{\sigma}(\tau_2)\hat{a}^\dagger_{\mathrm{in}}(t_{2})\ket{0} \delta(t_{1}-\tau_{1}) \Theta(t_2-t_1) \Theta(\tau_2-\tau_1)\nonumber \\
    & -\sqrt{2\gamma}\bra{0}\hat{\sigma}(\tau_2)\hat{a}^\dagger_{\mathrm{in}}(t_{1})\ket{0} \delta(t_{2}-\tau_{1}) \Theta(t_2-t_1) \Theta(\tau_2-\tau_1)\nonumber \\ & + \mathrm{G}_\mathrm{bb}
\end{align}

This can be simplified using the input-output relation $\hat{a}_\mathrm{in}^\dagger(t) = \hat{a}_\mathrm{out}^\dagger(t)+\sqrt{2\gamma}\hat{\sigma}^\dagger(t)$ and the quantum causality condition \eqref{eq:appendix_causality}(b) to obtain:

\begin{align}
    & \implies G_{\mathrm{ab}} \nonumber \\  & = -2\gamma e^{-2\gamma(\tau_2-t_2)} \Theta(\tau_2-t_2) \Theta(t_2-\tau_1)\delta(t_1-\tau_1) \nonumber \\
    & -2\gamma e^{-2\gamma(\tau_2-t_1)} \Theta(\tau_1-t_1) \delta(t_2-\tau_1) \nonumber \\ & + \mathrm{G}_\mathrm{bb}
\label{eq:appendix_Grt}
\end{align}

Again, using the input-output relations and the causality conditions, the following scattering amplitude can be obtained:

\begin{align}
    \psi_{\mathrm{ab}}(\tau_1,\tau_2) & = - 2\gamma e^{-2\gamma \tau_2}\int_{-\infty}^{\tau_1} \d t_1 ~ e^{2\gamma t_1}\xi(t_1,\tau_1) \nonumber \\ 
    & - 2\gamma e^{-2\gamma \tau_2} \int_{\tau_1}^{\tau_2} \d t_2 ~ e^{2\gamma t_2}\xi(\tau_1,t_2) \nonumber \\ 
    & + \psi_{\mathrm{bb}} (\tau_1,\tau_2)
\end{align}

The only remaining possibility is the transmission of both incident photons to the output port $\hat{a}_\mathrm{out}.$ This corresponds to $\mathrm{G}_\mathrm{aa}$, given by:

\begin{align}
    G_\mathrm{aa} & = \bra{0}\hat{a}_\mathrm{out}(\tau_2)\hat{a}_\mathrm{out}(\tau_1)\hat{a}^\dagger_\mathrm{in}(t_1) \hat{a}^\dagger_\mathrm{in}(t_2)\ket{0} \nonumber \\
    & = \bra{0}\hat{a}_\mathrm{in}(\tau_2)\hat{a}_\mathrm{in}(\tau_1)\hat{a}^\dagger_\mathrm{in}(t_1) \hat{a}^\dagger_\mathrm{in}(t_2)\ket{0} \nonumber \\ 
    & - \sqrt{2\gamma}\bra{0}\hat{a}_\mathrm{in}(\tau_2)\hat{\sigma}(\tau_1)\hat{a}^\dagger_\mathrm{in}(t_1) \hat{a}^\dagger_\mathrm{in}(t_2)\ket{0} \nonumber \\ 
    & - \sqrt{2\gamma}\bra{0}\hat{\sigma}(\tau_2)\hat{a}_\mathrm{in}(\tau_1)\hat{a}^\dagger_\mathrm{in}(t_1) \hat{a}^\dagger_\mathrm{in}(t_2)\ket{0} \nonumber \\ 
    & + 2\gamma\bra{0}\hat{\sigma}(\tau_2)\hat{\sigma}(\tau_1)\hat{a}^\dagger_\mathrm{in}(t_1) \hat{a}^\dagger_\mathrm{in}(t_2)\ket{0}
\end{align}

Using eqns.~\eqref{eq:appendix_Grr},~\eqref{eq:appendix_Gtr} and~\eqref{eq:appendix_Grt}, $G_\mathrm{aa}$ can  be expressed in terms of $G_\mathrm{ab}$, $G_\mathrm{ba}$ and $G_\mathrm{bb}$:
\begin{equation}
\implies G_\mathrm{aa} = \delta(\tau_2-t_2)\delta(\tau_1-t_1)+G_\mathrm{ab}+G_\mathrm{ba}-G_\mathrm{bb}    
\end{equation}

Similarly, as done above, we have the scattering amplitude: 

\begin{equation}
    \psi_{\mathrm{aa}} = \xi(\tau_{1}, \tau_{2}) + \psi_{\mathrm{ab}}(\tau_{1}, \tau_{2}) + \psi_{\mathrm{ba}}(\tau_{1}, \tau_{2}) - \psi_{\mathrm{bb}}(\tau_{1}, \tau_{2})
\end{equation}

\section{Derivation of splitting efficiency}

The splitting efficiency can be calculated from the scattering amplitudes derived in the previous section. To do this, we first calculate the scattering amplitudes at the output of the Mach-Zehnder Interferometer at ports $\hat{c}_{\mathrm{out}}$ and $\hat{d}_{\mathrm{out}}$. We provide analytical solutions for these scattering amplitudes $\psi_\mathrm{cc}$, $\psi_\mathrm{cd}$, $\psi_\mathrm{dc}$, $\psi_\mathrm{dd}$ below. The probability density of splitting is then given by $\rho_{s}=|\psi_\mathrm{cd}|^2+|\psi_\mathrm{dc}|^2$.

To calculate the scattering amplitudes at the output of the interferometer, we use the input-output relations of the Mach-Zehnder Interferometer from eqn.~\eqref{eq:MZI} to express them in terms of the scattering amplitudes at the output of the two-level emitter. We illustrate this procedure by showing the steps for calculating $\psi_\mathrm{cd}$ explicitly:
\begin{widetext}

\begin{align}
    \psi_\mathrm{cd} & = \bra{0}\hat{d}_\mathrm{out}(\tau_2) \hat{c}_\mathrm{out}(\tau_1)\ket{0} \nonumber \\ 
    & = \bra{0}\left( e^{i\phi}\cos \left( \frac{\theta}{2} \right) \hat{a}_\mathrm{out}(\tau_2)-\sin\left( \frac{\theta}{2} \right)\hat{b}_\mathrm{out}(\tau_2) \right) \left( e^{i\phi} \sin\left( \frac{\theta}{2} \right) \hat{a}_\mathrm{out}(\tau_1)+ \cos\left( \frac{\theta}{2} \right) \hat{b}_\mathrm{out}(\tau_1)\right) \ket{0} \nonumber \\
    & = - \sin\left( \frac{\theta}{2} \right) \cos\left( \frac{\theta}{2} \right)\psi_\mathrm{bb}+e^{i\phi} \cos^2\left( \frac{\theta}{2} \right)\psi_\mathrm{ba}-e^{i\phi} \sin^2\left( \frac{\theta}{2} \right)\psi_\mathrm{ab}+e^{2i\phi}\sin\left( \frac{\theta}{2} \right) \cos\left( \frac{\theta}{2} \right)\psi_\mathrm{aa}
\end{align}
\end{widetext}

where the first step follows from the definition of $\psi_\mathrm{cd}$. The second step follows from the input-output relations of the interferometer (see eq.~\eqref{eq:MZI}). The third step uses the definition of the scattering amplitudes at the output of the two-level emitter. Using the above procedure, we get:

\begin{widetext}
    \begin{equation}
    \psi_\mathrm{dc}=-\sin \left( \frac{\theta}{2} \right) \cos\left( \frac{\theta}{2} \right)\psi_\mathrm{bb} - e^{i\phi} \sin^{2}\left( \frac{\theta}{2} \right)\psi_\mathrm{ba} + e^{i\phi} \cos^2\left( \frac{\theta}{2} \right)\psi_\mathrm{ab}+e^{2i\phi}\sin\left( \frac{\theta}{2} \right) \cos\left( \frac{\theta}{2} \right)\psi_\mathrm{aa}
    \end{equation}
    \begin{equation}
    \psi_\mathrm{cc}= \cos^2\left( \frac{\theta}{2} \right)\psi_\mathrm{bb}+e^{i\phi} \sin\left( \frac{\theta}{2} \right) \cos\left( \frac{\theta}{2} \right)\psi_\mathrm{ba}+e^{i\phi}\sin\left( \frac{\theta}{2} \right) \cos\left( \frac{\theta}{2} \right)\psi_\mathrm{ab}+e^{2i\phi} \sin^2\left( \frac{\theta}{2} \right)\psi_\mathrm{aa}
    \end{equation}
     \begin{equation}
    \psi_\mathrm{dd} = \sin^2\left( \frac{\theta}{2} \right)\psi_\mathrm{bb}-e^{i\phi} \sin\left( \frac{\theta}{2} \right)\cos\left( \frac{\theta}{2} \right)\psi_\mathrm{ba}-e^{i\phi}\sin\left( \frac{\theta}{2} \right)\cos\left( \frac{\theta}{2} \right)\psi_\mathrm{ab}+e^{2i\phi}\cos^2\left( \frac{\theta}{2} \right)\psi_\mathrm{aa}
    \end{equation}
\end{widetext}

We note that $|\psi_\mathrm{cc}|^2+|\psi_\mathrm{cd}|^2+|\psi_\mathrm{dc}|^2+|\psi_\mathrm{dd}|^2=|\psi_\mathrm{bb}|^2+|\psi_\mathrm{ba}|^2+|\psi_\mathrm{ab}|^2+|\psi_\mathrm{aa}|^2.$ Since the right hand side of this equation integrates to 1 over the output times $\tau_1$ and $\tau_2$, so does the left hand side, ensuring proper normalization of the output photon wavefunction. This preservation of probabilities is ensured by the unitarity of the MZI transformation of eq.~\eqref{eq:MZI}. 

We obtain the probability density of the two input photons being split to different output modes of the MZI as $\rho_s=|\psi_\mathrm{cd}|^2+|\psi_\mathrm{dc}|^2$, which gives:

\begin{widetext}
\begin{multline}
    \rho_\mathrm{s}= \frac{1}{4} (2 \psi_\mathrm{bb} \left(-\sin (2 \theta ) \cos (\phi ) (\psi_\mathrm{ba}+\psi_\mathrm{ab})-8 \sin ^2(\theta/2 ) \cos ^2(\theta/2 ) \cos (2 \phi ) \psi_\mathrm{aa}\right)+2 \sin ^2(\theta ) \psi_\mathrm{bb}^2+2 \psi_\mathrm{ba} ((\cos (2\theta )-1) \psi_\mathrm{ab}+\\\sin (2\theta ) \cos (\phi ) \psi_\mathrm{aa})+(\cos (2\theta )+3) \psi_\mathrm{ba}^2+2 \sin (2\theta ) \cos (\phi ) \psi_\mathrm{ab} \psi_\mathrm{aa}+(\cos (2\theta )+3) \psi_\mathrm{ab}^2+2 \sin ^2(\theta/2 ) \psi_\mathrm{aa}^2)
    \label{eq:appendix_rhos}
\end{multline}

\end{widetext}

The routing efficiency is given by the integral of $\rho_s$ over t and $\tau.$ 

\section{Analytical Solutions for Splitting Efficiency \label{sec:appendix_anasol}}

In the main text, we presented results for the splitting efficiencies of Gaussian and Exponential pulse profiles. In the case of Gaussian pulses, the calculation of splitting efficiency has to be performed numerically. However, for exponential profiles, the splitting efficiency can be calculated analytically. In this section, we present the analytical results for the splitting efficiency in the case of uncorrelated and entangled exponential inputs.

The input state of two uncorrelated photons with an exponential pulse profile is given by $\xi(t_1,t_2)=\sqrt{2}\xi(t_1)\xi(t_2)$ with $\xi(t) = \sqrt{2\kappa}e^{-\kappa t}$. Plugging this input state into eqns.~\eqref{eq:scat_amps} yields the following expressions for the two photon wavefunctions at the outputs $a_{\mathrm{out}}$ and $b_{\mathrm{out}}$ of the two-level atom:

\begin{widetext}
\begin{subequations}
\begin{equation}
   \psi_{\mathrm{bb}}(\tau_1,\tau_2)= -\frac{8 \sqrt{2} \gamma ^2 \kappa  \left(e^{2 \gamma  \tau_1}-e^{\kappa  \tau_1}\right) e^{-2 \tau_1 (\gamma +\kappa )-\tau_2 (2 \gamma +\kappa )} \left(e^{2 \gamma  \tau_1+\kappa  \tau_2}-e^{\kappa  \tau_1+2 \gamma  \tau_2}\right)}{(\kappa -2 \gamma )^2}
\end{equation}

\begin{equation}
    \psi_{\mathrm{ba}}(\tau_1,\tau_2)=-\frac{4 \sqrt{2} \gamma  \kappa  \left(e^{2 \gamma  \tau_1}-e^{\kappa  \tau_1}\right) e^{-2 \tau_1 (\gamma +\kappa )-\tau_2 (2 \gamma +\kappa )} \left(2 \gamma  e^{2 \gamma  \tau_1+\kappa  \tau_2}-\kappa  e^{\kappa  \tau_1+2 \gamma  \tau_2}\right)}{(\kappa -2 \gamma )^2}
\end{equation}

\begin{equation}
    \psi_{\mathrm{ab}}(\tau_1,\tau_2)= \frac{4 \sqrt{2} \gamma  \kappa  e^{-2 \tau_1 (\gamma +\kappa )-\tau_2 (4 \gamma +\kappa )} \left(-2 \gamma  e^{2 \kappa  \tau_1+4 \gamma  \tau_2}-2 \gamma  e^{4 \gamma  \tau_1+2 \gamma  \tau_2+\kappa  \tau_2}+(4 \gamma -\kappa ) e^{(2 \gamma +\kappa ) (\tau_1+\tau_2)}+\kappa  e^{2 \gamma  \tau_1+\kappa  \tau_1+4 \gamma  \tau_2}\right)}{(\kappa -2 \gamma )^2}
    \end{equation}
    
\begin{equation}
    \psi_{\mathrm{aa}} = \xi(\tau_{1}, \tau_{2}) + \psi_{\mathrm{ab}}(\tau_{1}, \tau_{2}) + \psi_{\mathrm{ba}}(\tau_{1}, \tau_{2}) - \psi_{\mathrm{bb}}(\tau_{1}, \tau_{2})
\end{equation}
\label{eq:scat_amps_uncorr_exp}
\end{subequations}
\end{widetext}

Using these expressions in eqn.~\eqref{eq:appendix_rhos}, we can calculate probability density that the two input photons are routed to different output ports $c_\mathrm{out}$ and $d_\mathrm{out}$ of the interferometer. Integrating the resulting expression over $\tau_1$ and $\tau_2$ gives the following splitting efficiency:
\begin{widetext}
\begin{equation}
 P_{\mathrm{S}}=   \frac{(\kappa  (\kappa  (10 - 3 \kappa )+20)-8) \cos (2\theta)+16 \kappa  \sin ^2( \theta ) \cos (2 \phi )+32 \kappa  \sin (2\theta ) \cos (\phi )+\kappa  (\kappa  (3 \kappa +38)+44)+8}{4 (\kappa +2)^2 (3 \kappa +2)}
\end{equation}
\end{widetext}

Here, we set the atomic bandwidth $\gamma=1$. Note that this corresponds to expressing the pulse bandwidth $\kappa$ in the units of the atomic bandwidth $\gamma.$

The input state of two entangled photons with an exponential pulse profile is given by $\xi(t_1,t_2)=2 \sqrt{\kappa  \delta }~e^{-\kappa  t_1} e^ {-\delta  (t_2 -t_1)}$. We follow the same steps as in the previous section to obtain the two-photon wavefunction at the output of the two level atom. The resulting splitting efficiency in the stationary limit is given by:

\begin{widetext}
\begin{equation}
   P_{\mathrm{S}}= \frac{-(\delta  ((\delta -10) \delta -12)+8) \cos (2\theta )+16 \delta  \sin ^2( \theta ) \cos (2 \phi )+32 \delta  \sin (2 \theta ) \cos (\phi )+\delta  (\delta  (\delta +22)+20)+8}{4 (\delta +2)^3}
\end{equation}    
\end{widetext}

where the stationary limit corresponds to taking the limit $\kappa \to 0.$ Also note that we have set the atomic bandwidth $\gamma=1$, which merely corresponds to expressing the pulse bandwidth $\delta$ in terms of the atomic bandwidth.


\end{document}